\documentclass[useAMS,usenatbib, nofootinbib,11pt]{elsarticle}
\usepackage{amsmath,amsfonts,bbm, graphicx,color}
\usepackage{aas_macros}
\usepackage{url}
\usepackage{multirow}
\usepackage{float}
\usepackage{amsmath}
\usepackage{algorithm}
\usepackage[noend]{algpseudocode}
\makeatletter
\def\BState{\State\hskip-\ALG@thistlm}
\makeatother
\algdef{SE}[SUBALG]{Indent}{EndIndent}{}{\algorithmicend\ }%
\algtext*{Indent}
\algtext*{EndIndent}
\def\be{\begin{equation}}
\def\ee{\end{equation}}
\def\ba{\begin{eqnarray}}
\def\ea{\end{eqnarray}}

\begin{document}

\begin{frontmatter}

 \title{{\it astroABC}: An Approximate Bayesian Computation Sequential Monte Carlo sampler for cosmological parameter estimation} 


\author[1,2]{Elise Jennings}
\author[3]{Maeve Madigan}

\address[1]{Center for Particle Astrophysics, Fermi National Accelerator Laboratory MS209, P.O. Box 500, Kirk Rd. \& Pine St., Batavia, IL 60510-0500}
\address[2]{Kavli Institute for Cosmological Physics, Enrico Fermi Institute, University of Chicago, Chicago, IL 60637}
\address[3]{Department of Theoretical Physics, University of Dublin, Trinity College, Dublin, Ireland}

\begin{abstract}
Given the complexity of modern cosmological parameter inference where we are faced with non-Gaussian data and noise, correlated systematics and multi-probe correlated data sets, the Approximate Bayesian Computation (ABC) method is a promising alternative to traditional Markov Chain Monte Carlo approaches in the case where the Likelihood is intractable or unknown.
The ABC method is called ``Likelihood free" as it avoids
 explicit evaluation of the Likelihood by using a forward model simulation of the data which can include systematics.
We introduce {\it astroABC}, an open source ABC Sequential Monte Carlo (SMC) sampler for parameter estimation.
A key challenge in astrophysics is the efficient use of large multi-probe datasets to constrain high dimensional, possibly correlated parameter spaces. With this in mind {\it astroABC} allows for massive parallelization using MPI, a framework that handles spawning of processes across multiple nodes.  
A key new feature of {\it astroABC} is the ability to create MPI groups with different communicators, one for the sampler and several others for the forward model simulation, which speeds up sampling time considerably. 
For smaller jobs the Python multiprocessing option is also available.
Other key features of this new sampler include:
a Sequential Monte Carlo sampler;
a method for iteratively adapting tolerance levels;
 local covariance estimate using scikit-learn's  KDTree;
modules for specifying optimal covariance matrix for a component-wise  or multivariate normal perturbation kernel  and a weighted covariance metric;
restart files output frequently so an interrupted sampling run can be resumed at any iteration;
output and restart files are backed up every iteration;
user defined distance metric and simulation methods;
a module for specifying heterogeneous parameter priors including non-standard prior PDFs;
a module for specifying a constant, linear, log or exponential tolerance level;
well-documented examples and sample scripts. 
This code is hosted online at https://github.com/EliseJ/astroABC
\end{abstract}

\end{frontmatter}

\section{Introduction}

Given the size and complexity of modern cosmological data, Bayesian methods are now standard analysis procedures.
Bayesian inference allows us to efficiently combine datasets
from different probes, to update or incorporate prior information into parameter inference and to carry out model selection or comparison with Bayesian Evidence.
Cosmology is the latest discipline to embrace Approximate Bayesian methods, a development
driven by both the complexity of the data and covariance matrix estimation, together with the availability of new algorithms for running fast simulations of mock astronomical datasets \citep{2013ApJ...764..116W, 2012MNRAS.425...44C, 2015A&A...583A..70L}.
ABC and so called ``Likelihood free" Markov chain Monte Carlo (MCMC) techniques are popular methods
for tackling parameter inference in scenarios where the Likelihood (probability of the data given the parameters) is intractable or unknown \citep[see e.g.][]{TVZ}.
These methods are called ``Likelihood free" as they avoid
 explicit evaluation of the Likelihood by using a forward model simulation of the data.
 In Likelihood free MCMC techniques the acceptance probability  is computed using a metric between the data and the simulation of the 
 data without using a likelihood function.
 ABC methods aim to simulate samples from the parameter posterior distribution (probability of the parameters given the data) directly.
In MCMC approaches the target distribution is the posterior probability distribution function (pdf) of interest. In practice our estimate of this pdf is approximate due to finite sampling time, resulting in a correlated chain which we hope has converged.
ABC methods are also approximate in the sense that samples are generated from trial distributions which we hope are close to the real posterior of interest.
In this paper we present {\it astroABC}, a new open source Python ABC Sequential Monte Carlo sampler which allows for massive parallelization using MPI.

The standard in cosmological parameter estimation is to adopt a Bayesian approach, where a Likelihood function, together with a prior pdf for the parameters of interest, are sampled over using an MCMC to simulate from the posterior distribution. There are many public parameter
estimation codes available to the astrophysics community which focus on MCMC methods for analyzing complex cosmological datasets, as well
as calculating the physical analytical models and covariances which are needed in the Likelihood \citep[e.g.][]{2002PhRvD..66j3511L, 2014MNRAS.440.1379E,2015A&C....12...45Z}. Looking to the future of cosmological parameter inference, most of the statistical constraining power will come from combining datasets from multiple different probes \citep{ 2013PhR...530...87W,2014MNRAS.440.1379E, 2016arXiv160701014N}. Evaluating the Likelihood for combined probes is a non-trivial task as complex physical data is unlikely to have a simple multi-Gaussian or analytical form. Accounting for modeling  and instrumental systematics, and significant correlations between the parameters of interest and nuisance parameters in either the covariance matrix or Likelihood can be a daunting task \citep{2013PhRvD..88f3537D, 2013JCAP...11..009M}. 
ABC replaces the calculation of the Likelihood function with a simulation that produces a mock data set which can be compared to the observed data.
ABC methods are a promising alternative as a forward simulated model for the data includes systematics and correlations self consistently. 

There are several implementations of ABC methods already available in the R\footnote{https://www.r-project.org/} language, available
as packages from CRAN e.g  easyabc \footnote{http://easyabc.r-forge.r-project.org/} and abc \citep{2011arXiv1106.2793C}; as well as Python
packages e.g abcpmc \citep{2015ascl.soft04014A} and cosmoabc \citep{2015A&C....13....1I}. 
{\it astroABC} has several innovative new factors which make this software of value to the astrophysics community.

A key advancement in {\it astroABC} is the facility for massive parallelization using MPI, a framework that handles spawning of jobs across multiple nodes and provides a mechanism for communication between the processors. 
Allowing for MPI communication is a crucial speed enhancement which, to our knowledge, is not 
available in any of the Python abc packages currently in use in astronomy.  
For smaller jobs the Python multiprocessing option is also available which can spawn multiple processes but which are still bound within a single node. Other innovative factors include: several end user options are available to specify how the covariance, kernel and tolerance levels are calculated for the particles; different priors (as well as non-standard prior distributions) can be specified for each parameter; restart files are written every iteration to allow for e.g. interrupted sampling runs or time limits on batch queue systems; output and restart files are backed up every iteration. 
As many in the astrophysics community are not familiar with the statistics package {\sc R},  {\it astroABC} is available as a set of Python modules which are well documented and easily modified if necessary in a specific application.
In addition  {\it astroABC}  will be made available independently on github and as part of the {\it CosmoSIS} parameter estimation code \citep{2015A&C....12...45Z}.
{\it astroABC} is packaged for distribution using {\sc PYPI}\footnote{https://pypi.python.org/pypi}; is publicly available on github\footnote{https://github.com/EliseJ/astroABC} and synced with Travis continuous integration \footnote{https://travis-ci.org/} for testing new builds or commits. It is well documented with class/method doc strings as well as tutorial and example pages hosted on the github wiki.

\section {Bayesian Analysis and Markov Chain Monte Carlo}
In this section we give a brief background on Bayesian inference and traditional MCMC methods which will be useful when comparing with 
Approximate Bayesian Computation.

\subsection{Bayesian Inference}

The fundamental problem in Bayesian statistics is the computation of posterior distributions.
We are interested in estimating the posterior pdf for some underlying parameters, ${\bf \theta}$ of a model, $M$, given some data and prior 
information about those parameters. Bayes's Theorem allows us to write this posterior distribution in terms of the Likelihood for the data, $\mathcal{L}({ D}|M({\bf \theta}))$, and the prior distribution, $\pi({ \theta})$, as
\ba
P({\bf \theta}|{D}) =  \frac{\mathcal{L}({D}|M({\theta})) \pi(\bf{\theta})}{\int \mathcal{L}({{D}}|M({\bf \theta})) \pi(\bf{\theta}) {\rm d}{\theta}}
\ea
where the denominator is referred to as the Bayesian Evidence or marginal Likelihood; and the integral runs over all possible parameter values.
The prior probability represents our state of knowledge of the data and may incorporate results from previous datasets; restrict the range for 
physical parameters e.g. masses must be positive; or may be un-informative with little restriction. 
The choice of Likelihood for many cosmological analysis is a single or multivariate Gaussian where the mean is evaluated using some physical model and the covariance matrix is measured or estimated either analytically or numerically. 
In this framework the accuracy of the parameter estimation will depend heavily on our choice for the Likelihood, as well as the accuracy of the physical model for the data, and how well parameter covariance and correlated systematics are described in the covariance matrix \citep[see e.g.][]{2015A&C....12...45Z,2014MNRAS.440.1379E}.
For a review of probability, parameter inference and  numerical techniques such as MCMC methods please see e.g. \citep{2008ConPh..49...71T, 2009arXiv0906.0664H, Jaynes}.

\subsection{Markov Chain Monte Carlo techniques}

MCMC techniques are an efficient way to simulate from the posterior pdf when analytical solutions do not exist or are intractable.
An MCMC algorithm constructs a sequence of points in parameter
space, referred to as an MCMC chain, 
which is a discrete time stochastic process where each event in the chain is generated from the
Markov assumption that the probability
of the $(i+ 1)^{th}$ element in the chain only depends on the value of the $i^{th}$
element. 
Markov Chains are called ``memory-less'' because of this assumption.
A key property of Markov chains is that under certain conditions the distribution of the chain evolves to a
stationary or target state  independently of its initial starting point.
If our target distribution is the posterior pdf then we want the unique limiting
distribution for the Markov Chain to be simulated draws from this posterior distribution.
Many MCMC algorithms exist, including the Metropolis-Hastings algorithm \citep{MCMC}, Gibbs sampling, Hamiltonian Monte Carlo, 
importance sampling and ensemble sampling \citep[see e.g.][]{GoodmaneWeare}. 
Each method relies on a proposal distribution (which may have separate parameters which need to be tuned) to advance events in the chain from the starting distribution towards the target pdf.
Once the chain has converged the density of points in the chain is proportional to the posterior pdf.
If the Likelihood and model are correct then MCMC will lead to a posterior pdf that converges to a common
inference on the model parameters.

\section{ Approximate Bayesian Computation  \label{section:abc_smc}}
In Section \ref{sec:abc_intro} we describe ABC and motivate its use for cosmological parameter estimation, in Section \ref{sec:abc_algo}
we describe a general ABC Sequential Monte Carlo algorithm and in Section \ref{sec:metric} we discuss the ABC distance metric and sufficiency conditions on summary statistics.

\subsection{ABC: parameter inference without Likelihood assumptions \label{sec:abc_intro}}

In traditional MCMC approaches the Likelihood used (most often a simple multi-Gaussian) is a key assumption in the method. With incomplete analytical expressions for the Likelihood or computational restrictions on how accurately we can estimate the covariance matrix, this assumed pdf will be incorrect, leading to biased parameter constraints.
For example in supernova analysis studies, it is well known that selection effects (e.g magnitude or color cuts) imposed on the data sample are not easily accounted for analytically in the Likelihood and can lead to biased population parameters from MCMC sampling; or in analyzing the number counts of clusters of galaxies, analytical models in the Likelihood are often calibrated  using a particular definition of a cluster which may not be applicable for the data, or using an N-body simulation in a fixed
cosmological model. 
ABC methods aim to simulate samples directly from the parameter posterior distribution of interest without assuming a particular form for the Likelihood.

\subsection{ ABC algorithms \label{sec:abc_algo}}

The simplest ABC algorithm is rejection sampling. 
Given a set of parameters, $\theta$, with associated priors, $\pi(\theta)$ and a forward simulated model for the data,
$\pi(D|\theta)$, 
we can simulate from the posterior distribution, $P(\theta|D)$, by first drawing sample parameters
$\theta^* \sim \pi(\theta)$, 
then simulating a dataset with these parameters 
$D^* \sim \pi(D|\theta^*)$.
In a  rejection sampling algorithm, we reject $D^*$  unless it matches the true data, $D$.
For a discrete data set, which is a single random realization from a continuous pdf, this algorithm would not be practical as many simulated samples  would be rejected until an exact match is found.
In practice we make an approximation and accept simulated datasets which are ``close" to the true data. This 
notion of simulating a mock dataset which is close to the observed data
introduces the ideas of a distance metric and tolerance level in ABC. We accept proposed parameters $\theta^*$, if 
$\rho(D^* - D) <\epsilon$
where $\rho$ is the distance metric, which could be e.g. the Euclidean norm $||D^* - D||$,  and $\epsilon$ is a tolerance threshold. This procedure produces samples from the pdf
$P(\theta | \rho(D^*-D)<\epsilon)$,
which will be a good approximation of the true posterior if $\epsilon$ is small.

Rather than drawing candidates $\theta^*$, one at a time, we can
speed up the ABC algorithm by working with large
pools of candidates, called particles, simultaneously. 
At each stage of the algorithm the particles are perturbed and filtered using the distance metric, and eventually
this pool of particles move closer and closer to simulating from the desired posterior distribution.
This approach is known as Sequential Monte Carlo or Particle Monte Carlo sampling and the algorithm is presented in 
Algorithm 
 \ref{algo_abc} \citep[see e.g.][]{2008arXiv0805.2256B,2009arXiv0901.1925T, 2010arXiv1001.2058S}.
    
Different ABC SMC algorithms can be distinguished
by how sampling weights are assigned to the particles in the pool.
In order to filter and perturb the particles we need a transition kernel. The transition
kernel serves the same purpose as the proposal distribution in a standard
MCMC algorithm. The transition kernel specifies the distribution of a random variable that will
be added to each particle to move it around in the parameter space.
At iteration $t$, the ABC SMC algorithm proposes parameters from the following
\ba
q_t{\theta} = \begin{cases} \pi(\theta), \qquad \qquad\qquad \qquad \qquad \qquad {\rm if} \, t=0 \\
\sum^{N}_{j=1} w_{j,t-1} \mathcal{K}(\theta_{j,t-1}| \theta_{i,t}, \Sigma_{t-1}) , \qquad  {\rm otherwise}
\end{cases}
\ea

where $w_{j,t-1}$ are the weights for particle $j$ at iteration $t-1$ and $\Sigma_{t-1}$ is the covariance amongst the particles at $t-1$.
This effectively filters out a particle from the previous weighted pool, then perturbs the result using the kernel ${\mathcal K}$.
The weighting scheme in ABC SMC  minimizes the
Kullback -- Leibler distance, a measure of the discrepancy between two
density functions.  Minimizing the Kullback -- Leibler distance, between the desired posterior and
the proposal distribution,
maximizes the acceptance probability in the algorithm \citep{2011arXiv1106.6280F}.
For more details on the different choices of kernel as well as optimization techniques see e.g.\citep{2008arXiv0805.2256B, 2011arXiv1106.6280F}.
We list the kernel options available in {\it astroABC} in Section \ref{sec:runtime}.
In Algorithm 
 \ref{algo_abc} the ABC sampler is run for a maximum of $T$ iterations. In  {\it astroABC} the sampler runs until $T$ iterations has been achieved or until the minimum threshold specified is found. Both 
 the maximum number of iterations and the minimum threshold are user defined options and will depend on the required accuracy needed for the distance metric.

\begin{algorithm}
\label{alg:1}
\caption{ABC SMC algorithm for estimating the posterior distribution for parameters $\theta$ using $N$ particles, the prior distribution $\pi(\theta)$,
given data $D$ and a model for simulating the data $M(D|\theta)$. $\theta_{i,t}$ represents the parameter set for particle $i$ and iteration $t$.}\label{algo_abc}
\begin{algorithmic}[1]
\BState {Set the tolerance thresholds, $\epsilon_t$ for  $t=0\cdots T$ iterations.}
\Procedure{ ABC SMC LOOP}{} 
\BState{At iteration t=0:}
\For {$1\leq i \leq N$}
\While {$\rho(D,D^*) >\epsilon_0$}
\State Sample $\theta^*$ from prior $\theta^* \sim  \pi(\theta) $
\State Simulate mock data  $D^* \sim {\rm M}(D|\theta^*)$
\State Calculate distance metric $\rho(D,D^*)$ 
\EndWhile
\State Set $\theta_{i,0} \leftarrow \theta^*$
\State Set weights $w_{i,0} \leftarrow  1/N$
\EndFor
\State Set covariance $\Sigma_{0}^2 \leftarrow  2\Sigma(\theta_{1 :N,0})$
\BState{At iteration $t>0$:}
\For {$1<t<T$}
\For {$1\leq i \leq N$}
\While {$\rho(D,D^*) >\epsilon_t$}
\State Sample $\theta^*$ from previous iteration. $\theta^* \sim \theta_{1:N, t-1} $ with probabilities $w_{1:N,t-1}$
\State Perturb $\theta^*$ by sampling  $\theta^{**} \sim  \mathcal{N}(\theta^*, \Sigma^2_{t-1})$
\State Simulate mock data  $D^* \sim {\rm M}(D|\theta^{**})$
\State Calculate distance metric $\rho(D,D^*)$ 
\EndWhile
\State Set $\theta_{i,t} \leftarrow \theta^{**}$
\State Set weights $w_{i,t} \leftarrow  \frac{\pi(\theta_{i,t})}{\sum^{N}_{j=1} w_{j,t-1} \mathcal{K}(\theta_{j,t-1}| \theta_{i,t}, \Sigma_{t-1}) }$ using kernel $\mathcal{K}$
\EndFor
\State Set covariance $\Sigma_{t}^2$ using e.g. twice weighted empirical covariance 
\EndFor
\EndProcedure
\end{algorithmic}
\end{algorithm}

\subsection{The ABC metric and sufficient statistics \label{sec:metric}}

Using high-dimensional data can reduce the acceptance rate and reduce the efficiency of the ABC algorithm.
In many cases it may be simpler to work with some lower dimension summary statistic of the data, $S(D)$, e.g. the sample mean,
rather then the full dataset \citep{Marjoram}. In this case the chosen statistic needs to be a so-called {\it sufficient statistic} in that
any information about the parameter of interest which is contained in the data, is also contained in the summary statistic. More formally a statistic $S(D)$ is sufficient for $\theta$, if the distribution $P(D|S(D))$ does not depend on $\theta$.
This requirement ensures that in summarizing the data we have not thrown away constraining information about $\theta$.

The ABC method relies on some metric (a distance) to
compare the simulated data to the data that were observed.
It is common to use the weighted Euclidean distance,
\ba
\rho(S(D) - S(D^*)) = \left( \sum_i \left( \frac{S(D)_i - S(D^*)_i}{\sigma_i}\right)^2\right)^{1/2}
\ea
 between the  observed and simulated summary statistics as
a metric, where  $\sigma_i$ is the error on the $i^{\rm th}$ summary statistic \citep[see e.g.][]{Beaumont2002}. 
There are many choices for the ABC distance metric, for example, the  weighted sum of absolute differences, or $L_1$ distance, $\sum_i w_i |S(D)_i - S(D^*)_i|$. Choosing a summary statistic and distance metric which are sensitive to the parameters of interest is a crucial step in parameter inference.
The success of ABC relies on the fact that if the distance metric is defined by way
of sufficient statistics, then the resulting approximation to the
posterior will be good as long as  $\rho(S(D)-S(D^*))$ is less than some small threshold.
 There is currently a lot of scope for research on practical methods for identifying approximately sufficient statistics, and for assessing the adequacy of the distance metric in astrophysical examples. We will discuss this further in the example presented in Section \ref{sec:example}.

\section { {\it astroABC}}
In Section \ref{subsec:intro_abc} we discuss the open source code {\it astroABC}. There are several options for choosing how the particle covariance in the perturbation
kernel are estimated and these are presented in Section \ref{sec:runtime}. Further details of runtime options are also given in 
Section \ref{sec:runtime}.

\subsection{The {\it astroABC} code \label{subsec:intro_abc}}

{\it astroABC} is an open source Python ABC SMC sampler which can be run in parallel using MPI (with Python's mpi4py\footnote{https://pypi.python.org/pypi/mpi4py} package) or multithreading  
(with Python's multiprocessing package). 
The challenge in modern cosmological parameter inference is the efficient use of large multi-probe datasets to constrain high dimensional
correlated parameter spaces. In many cases there are complicated systematics which effect our ability to extract the cosmological parameters of interest with the required accuracy. Currently the approach has been to run computationally demanding simulations, which naturally include systematic uncertainties and correlations, to estimate 
a mean model and covariance matrix  for use in MCMC sampling, after assuming some form for the likelihood. This approach can be flawed as the number of simulations used can be insufficient to capture the full covariance matrix or the simulations are run in a fixed cosmology, limiting the usefulnesss of any estimates extracted.
The ABC sampling method is an alternative approach where a full forward model simulation, which includes systematics and uncertainties, is run at every point in parameter space. The simulation outputs from the ABC sampler vary with cosmology and are no longer fixed to one cosmological model. Any correlations between parameters are naturally included in the forward model simulation consistently at every point in parameter space so there is no need to use a potentially inaccurate covariance matrix in the distance metric. To estimate the covariance matrix in current MCMC approaches it is common  to run many simulations on a user defined grid in parameter space which may be insufficient. The ABC algorithm in  {\it astroABC} has been optimized to efficiently estimate the posterior distribution by running simulations only at select points in parameter space. 

The main limitation to the ABC method is computational speed, both for running the accept/reject algorithm and  the simulation.
With this in mind {\it astroABC} allows for massive parallelization using MPI.  A crucial feature is the ability to create MPI groups with separate communicators so that {\it astroABC} runs on a single MPI pool
of nodes, while each node in the pool has access to a separate MPI group with which to launch a simulation.
For smaller jobs a multiprocessing (multithreaded) option is also available which can spawn multiple processes but which are still bound within a {\it single} node.
This code has been designed with maximal flexibility in parameter inference where
the posterior pdf for multiple parameters is to be estimated. 

In summary this code allows the user to:
\begin{itemize}
\item create MPI groups with separate communicators so both  {\it astroABC} and simulation launched by each particle can run in parallel
\item run smaller jobs using Python's multiprocessing
\item assign different  prior distributions to each parameter
\item define the sufficient statistics and distance metric used in the sampling
\item set at runtime one of a choice of tolerance thresholds, particle covariance estimators and perturbation kernels
\item  frequently backup output files and restart files in case a job is canceled due to time restrictions.
\end{itemize}

End users must supply a dataset (or summary statistic of the dataset), a method for simulating the dataset (or the chosen summary statistic) and a method for the calculating the distance metric given the summary statistics of the data and simulation. 


\subsection{{\it astroABC} runtime options \label{sec:runtime}}

The end user should provide a method which takes an input set of parameters and returns a 
simulated dataset (or summary statistic of the dataset) for those parameters.
In initializing the {\it astroABC} class a user can specify the following:
\begin{itemize}
\item the number of particles which will simultaneously generate samples from the posterior distribution
\item the number of iterations
\item the number of parameters to vary in the sampling procedure
\item  the relevant priors for each parameter (together with the hyper parameters for each distribution) which is saved as a dictionary with  callable scipy.stats pdf functions for use in the sampling procedure.
\item draw from any non-standard prior PDF which might be particularly useful to cosmologists wanting to include priors from previous datasets e.g. using results from the Planck chains \citep{2015arXiv150201582P}. This could be the Markov chain samples or the output from any sampling method which returns a finite set of parameter and probability values from the 1D posterior distribution for that parameter. 
\end{itemize}

In {\it astroABC} we implement an adaptive transition kernel which depends on the covariance of accepted particles from the previous iteration \citep{2008arXiv0805.2256B}.
For particle $i$ at iteration $t$, this is a Gaussian kernel with mean $\theta_{i,t}$  and a covariance which can be 
either a component wise perturbation with local diagonal variance or a multivariate perturbation based on the covariance, depending on the end user's selection. 
I.e. the end user can either use a diagonal covariance matrix (neglecting correlations between parameters) or the full matrix.
The following options are available  to calculate the entries in the covariance matrix:
\begin{itemize}
\item twice the empirical covariance amongst the particles \citep[see e.g][]{TVZ}
\item covariances which result in a kernel which minimizes the Kullback-Leibler divergence between the target distribution and
the distribution of the perturbed particles
\item local covariance estimate using scikit-learn's  KDTree method for nearest neighbours \citep{2011arXiv1106.6280F}
\item twice the weighted particle covariance matrix \citep{2008arXiv0805.2256B} 
\item a shrinkage covariance metric with the Ledoit-Wolf estimator \citep{shrinkage}.
\end{itemize}

The tolerance level in {\it astroABC} controls which of the proposed parameters are accepted given the distance metric. There are two considerations in choosing a tolerance level. If the tolerance is too high then too many proposed parameters are accepted and the prior distribution dominates the results e.g. if the tolerance level is infinity then we would just recover the prior distribution from {\it astroABC}. If the tolerance level is too low then the ABC accept/reject algorithm is very inefficient with many proposed points being rejected.
A compromise is to select a set of decreasing tolerance levels where for the initial iterations we accept points in parameter space which do not represent the data with high accuracy but as the algorithm progresses the tolerance level decreases and the parameters are sampling a high probability region of the posterior distribution.
Within the {\it astroABC} tolerance class there are many options for selecting the tolerance type. Users can select a linearly decreasing, log decreasing, exponentially decreasing, constant or iteratively adaptive tolerance threshold. For the iteratively adaptive option the user specifies  a certain quantile  e.g. 75$^{\rm th}$,  of the metric distance from the previous iteration. As such this decreasing tolerance depends on the particle positions in the previous iteration. All tolerances require an input maximum and minimum value and the selected tolerance type is  then implemented from a maximum value until either the minimum value is reached or the particles have reached the maximum number of iterations requested.

\section{MCMC vs ABC example \label{sec:example}}
In this section we present a simple example of parameter inference, with a simulated supernovae dataset, using both a standard MCMC sampler and {\it astroABC}.
This example represents a nontrivial case which has been specifically chosen to highlight potential biases in the MCMC approach which can be avoided using ABC.
Note this is not meant to show the best constraints we can get from ABC versus MCMC using supernova data. The toy example is meant to breakdown key elements and assumptions about the two methods with a simple physical example and is intended as a teaching exercise.

\subsection{Example supernovae analysis}

Using the $~400$ supernovae in the redshift range $0.5<z<1.0$.
The `true' cosmological parameters for the mock data are $\Omega_m = 0.3$, the matter density of the universe today, 
$\Omega_{\Lambda} = 0.7$, the dark energy density of the universe today, 
 $w_0 = -1.0$, the present value of the dark energy equation of state and  $h_0 = 0.7$,  the current Hubble parameter. 

For demonstration purposes we add
artificial noise to the data at every redshift by adding a random variable from a skewed normal distribution with fixed parameters:
location = -0.1, scale = 0.3  and
skew =  5.0.\footnote{http://scipy.github.io/devdocs/generated/scipy.stats.skewnorm.html} After this procedure the output will be referred to as our dataset.
We analyze this dataset using an MCMC and the {\it astroABC} sampler and fit for two cosmological parameters: $\Omega_m$ and $w_0$.

Adding this non-Gaussian random noise means that the final dataset has a non-Gaussian distribution at every redshift.
In  realistic datasets there may be several systematics which will have a similar effect e.g  zero-point offset systematics in supernova studies.
We assume that an analytic expression for the distribution of the final dataset is not available to us but can be forward modeled using a simulation.
For this simple example we assume a Gaussian distribution for the data in our MCMC sampler.
On the other hand using {\it astroABC} we are able to run the simulation at every proposed point in parameter space during sampling and generate a mock dataset which includes possible sources of noise.
During sampling we assume a flat cosmological model and wide Normal priors on $\Omega_m \sim \mathcal{N}(0.3,0.5)$ and $w_0\sim \mathcal{N}(-1,0.5)$ and no prior information from other probes.
\begin{figure}
\begin{center}
\includegraphics[height=3.in,width=4.5in]{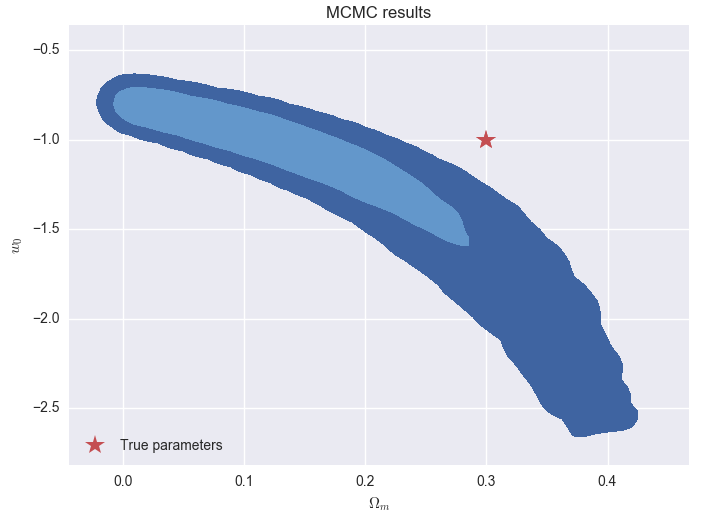}
\caption{Parameter constraints on $\Omega_m$ and $w_0$ from the MCMC sampler. The `true' parameter values used to generate the mock data are shown as a red star. The 1 and 2-$\sigma$ contours from the posterior distribution found using MCMC are shown as blue shaded regions.
}
\label{fig:mcmc}
\end{center}
\end{figure}

\subsection{MCMC results}

For our MCMC analysis we assume the following standard Likelihood:
\ba
\mathcal{L}(\mu_{\rm data} | \mu_{model}(z,\Omega_m,w_0)) \propto e^{- \sum_i \left( \frac{\mu^i_{\rm data} - \mu_{\rm model}(z^i, \Omega_m,w_0)}{2\sigma^i}\right)^2} \, ,
\ea
where $\mu^i_{\rm data}$ is the distance modulus for an individual supernova in the data, with associated error $\sigma_i$; and in a flat universe,
\ba
\mu_{\rm model}(z^i,\Omega_m,w_0)  \propto  5 {\rm log}_{10} \frac{c(1+z)}{h_0}\int_0^{z^i} {\rm d}z' \frac{1}{E(z')}
\label{eq:mu}
\ea
where
\ba
E(z) = \sqrt{\Omega_m (1+z)^3 + (1 - \Omega_m)e^{3 \int_0^z {\rm d ln}(1+z')[1+w(z')]} } 
\ea

The results of the estimated posterior distribution for $\Omega_m$ and $w_0$ from the  MCMC sampler are shown in Fig. \ref{fig:mcmc}. 
These results are from 28 chains each with 10,000 sample points.
The `true' parameters values used to generate the mock data are shown as a red star. The 1 and 2-$\sigma$ contours from the posterior distribution found using MCMC are shown as blue shaded regions.
The marginalized best fit values are $\Omega_m = 0.17 \pm 0.11$ and $w_0 =  -1.26 \pm 0.55$ which are quite discrepant with the `true' values of 
$\Omega_m = 0.3$ and $w_0 = -1.0$.
It is clear that  the simple Gaussian Likelihood assumption in this case, which neglects the effects of systematics yields biased cosmological constraints.

\subsection{ABC results}

As this is an example for demonstration purposes we are assuming that our noisy data can be simulated accurately and easily,
 but that an analytical expression for the likelihood is not available to us.
 
\begin{figure}
\begin{center}
\includegraphics[height=4in,width=5.5in]{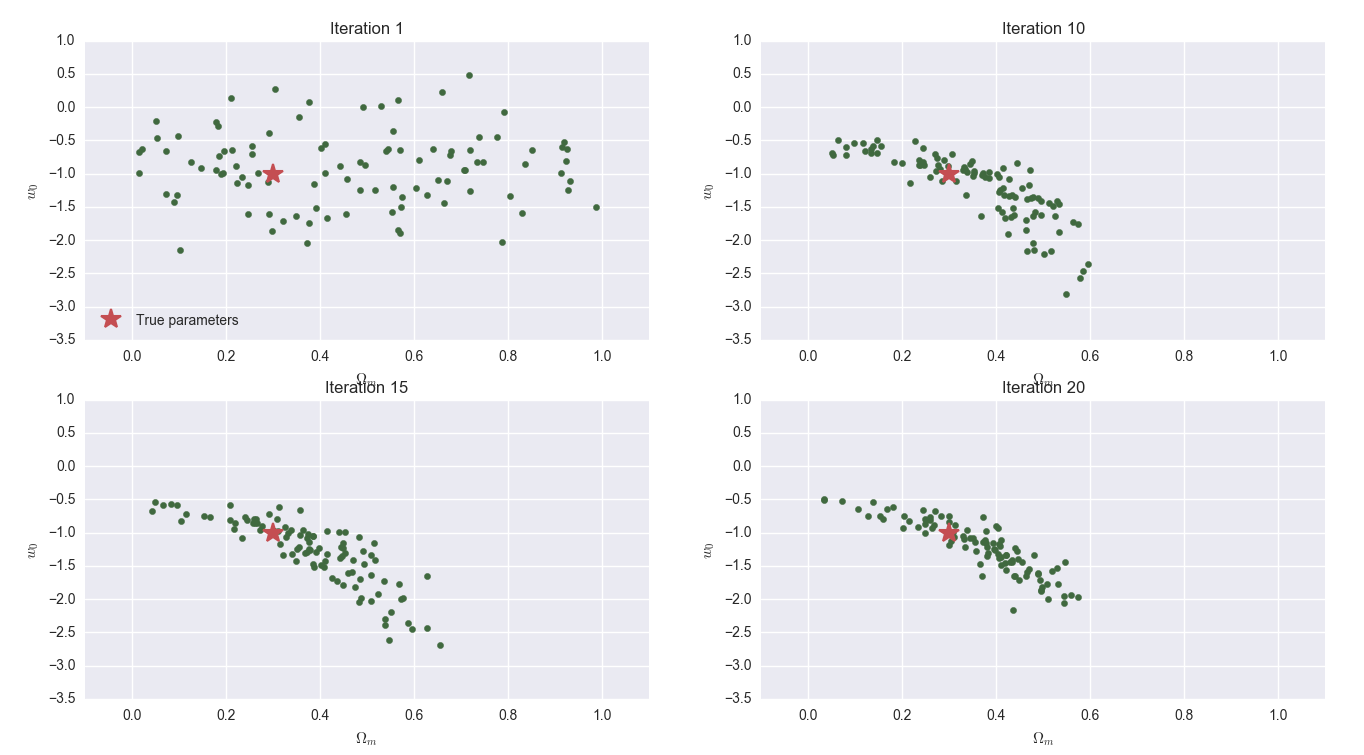}
\caption{The accepted $w_0$  and $\Omega_m$ parameter values 
at several iterations of the {\it astroABC} sampler. Each of the 100 particles is represented as a green dot. The `true' value of 
the parameters is shown as a red star in each panel.}
\label{fig:particles_iter}
\end{center}
\end{figure}

Using a forward model simulation we can account for 
non-Gaussian uncertainties in the data without explicitly knowing the likelihood. Our simulation in this case uses the model given in 
Eq. \ref{eq:mu} to draw Gaussian random variables to which we add non-Gaussian noise from a skew normal distribution.
Note the use of exactly the same model here in the two methods is to highlight the distinction between choice of {\it model}, and choice of {\it Likelihood} 
in inference techniques. Even though the physical model can be the same in both, an incorrect Likelihood assumption can bias results.
At every iteration we simulate a set of  supernovae at every point in a  two dimensional parameter space ($\{\Omega_m, w_0\}$).
With ABC it is also possible to parametrize the source of non-Gaussian noise  in the simulation and fit for e.g. the hyper parameters of the skew normal distribution also.
This simulated output is then compared with the dataset using a weighted Euclidean metric and an iteratively adaptive threshold.
We use 100 particles in the {\it astroABC} sampler and run until the error on the 1$\sigma$ contour for the parameters is $\sim 5$\%. 
There choice for the number of particles is based on trial and error runs to determine an optimum way to sample a multi dimensional parameter space efficiently.
Fig. \ref{fig:particles_iter} shows the progress of the ABC particles as the iteration number increases (and the tolerance level decreases). 
In each panel the `true' value of 
the parameters is shown as a red star.
It is clear that at iteration 0 the particles are well dispersed throughout the  prior range. As the tolerance threshold level decreases the particles converge towards the 
`true' values.

\begin{figure}
\begin{center}
\includegraphics[height=3in,width=4.5in]{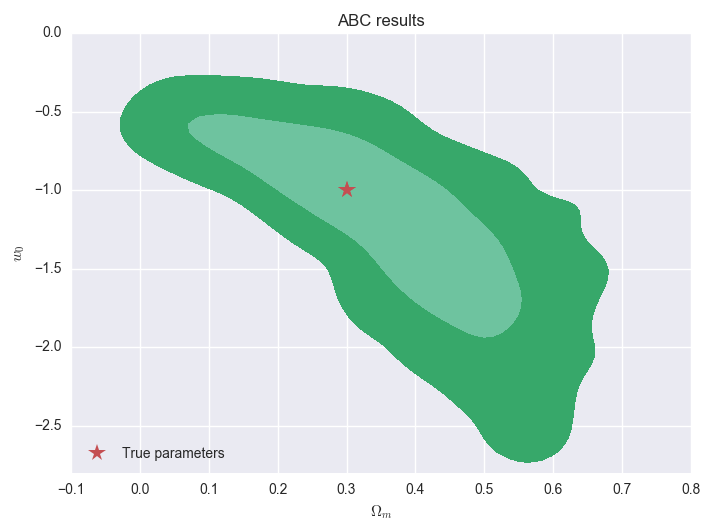}
\caption{Parameter constraints on $\Omega_m$ and $w_0$ from {\it astroABC}. The `true' parameter values used to generate the mock data are shown as a red star. The 1 and 2-$\sigma$ contours from the posterior distribution found using the {\it astroABC} sampler are shown as green shaded regions.
}
\label{fig:abc}
\end{center}
\end{figure}

In Fig. \ref{fig:abc} we show the 1 and 2-$\sigma$ contours from the joint posterior distribution of $\Omega_m$ and $w_0$ found using {\it astroABC}.
The marginalized best fit values are  $\Omega_m = 0.36 \pm 0.12$ and $w_0 = -1.22 \pm 0.4$ which recover the `true' underlying parameters used to create the dataset within the 1-$\sigma$ errors.
Using a forward simulation model, which can naturally incorporate systematic effects at every point in parameter space, avoids explicit calculation of the Likelihood and any parameter bias which was seen in the MCMC method.


\section{Discussion}

We have described {\it astroABC}, a new open source Python implementation of an ABC SMC sampler.
The ABC sampler can be used in parameter inference by simulating observations from posterior distributions when likelihoods are difficult or impossible to compute. Problems such as this arise frequently in cosmological applications, where it is often the case that a physical model for the data can be simulated rapidly, and includes systematic uncertainties, but is sufficiently complicated that explicit formulae for the Likelihood are not known. 
 The demand for alternative sampling methods in astronomy is increasing given the large correlated datasets we expect to analyze from future surveys. Current methods increasingly rely on simulations for error estimation and ABC sampling is a natural extension of this approach with the advantage that the Likelihood is not explicitly calculated. {\it astroABC} was designed to be as user friendly as possible while also accommodating  the computational demands of simulating future datasets. We hope that {\it astroABC} will be a useful resource for the astrophysics community with its facility for massive parallelization using MPI, as well as Python multiprocessing, and innovative factors such as a varied choice of covariance
and  kernel estimation, tolerance levels and priors.

\section{Acknowledgements}
We thank Rachel, Wolfe, Chad Schaffer, Rick Kessler, Dan Scolnic and Joe Zuntz for useful discussions.
EJ is grateful to I. Feely for inspiration and motivational kicks.
EJ is supported by Fermi Research Alliance, LLC under the U.S. Department of Energy under contract No. DE-AC02-07CH11359. 
Operated by Fermi Research Alliance, LLC under Contract No. De-AC02-07CH11359 with the United States Department of Energy.
MM  was supported by funding for the DES
Projects which been provided by the U.S. Department
of Energy, the U.S. National Science Foundation,
the Ministry of Science and Education
of Spain, the Science and Technology Facilities
Council of the United Kingdom, the Higher Education
Funding Council for England, the National
Center for Supercomputing Applications at
the University of Illinois at Urbana-Champaign,
the Kavli Institute of Cosmological Physics at the
University of Chicago, Financiadora de Estudos e
Projetos, Fundacao Carlos Chagas Filho de Amparo
a Pesquisa do Estado do Rio de Janeiro,
Conselho Nacional de Desenvolvimento Cientifico
e Tecnologico and the Ministerio da Ciencia e
Tecnologia, the Deutsche Forschungsgemeinschaft
and the Collaborating Institutions in the Dark Energy
Survey.
We are grateful for the support of
the University of Chicago Research Computing Center.

\bibliographystyle{mn2e}
\bibliography{thebibliography}

\end{document}